\documentclass{article}
\usepackage{amsmath}
\usepackage{graphicx}
\usepackage{amsfonts}
\usepackage{amsthm}
\usepackage{algorithm}
\usepackage{algorithmic}
\usepackage{times}
\usepackage{fancyhdr}
\newcommand{\argmax}{\operatornamewithlimits{argmax}}
\newcommand{\argmin}{\operatornamewithlimits{argmin}}

\newcommand{\PP}{\mathbb{P}}         
         
\newcommand{\EE}{\mathbb{E}}           
\newcommand{\II}{\mathbb{I}}           

\newcommand{\thma}{\begin{thm}}
\newcommand{\thmb}{\end{thm}}

\newtheorem{defi}{Definition}
\newtheorem{thm}{Theorem}

\newcommand{\enuma}{\begin{enumerate}}
\newcommand{\enumb}{\end{enumerate}}
\newcommand{\ena}{\begin{enumerate}}
\newcommand{\enb}{\end{enumerate}}

\newcommand{\itema}{\begin{itemize}}
\newcommand{\itemb}{\end{itemize}}
\newcommand{\ita}{\begin{itemize}}
\newcommand{\itb}{\end{itemize}}

\newcommand{\alignb}{\end{align}}

\newcommand{\proofa}{\begin{proof}}
\newcommand{\proofb}{\end{proof}}

\newcommand{\bla}{\begin{block}}
\newcommand{\blb}{\end{block}}

\newcommand{\seqb}{\end{equation*}}

\providecommand{\mc}[1]{\mathcal{#1}}

\providecommand{\mh}[1]{\hat{#1}}

\newcommand{\sig}{\sigma}

\newcommand{\eps}{\varepsilon}

\newcommand{\conv}{\rightarrow}

\newcommand{\defa}{\begin{defi}}
\newcommand{\defb}{\end{defi}}

\usepackage{amssymb}
\usepackage{hyperref}

\newcommand{\mB}{\mathcal{B}}
\newcommand{\mM}{\mathcal{M}}
\newcommand{\hL}{\widehat{L}}
\newcommand{\MeB}{\mM \overset{\varepsilon}{{\sim}}_{F} \mB}
\newcommand{\MsB}{\mM \overset{S}{\sim}_{F} \mB}
\newcommand{\from}{\colon}

\newtheorem{thex}{\emph{Gedankenexperiment}}
\begin{document}

\title{Are mental properties supervenient on brain properties?}
\author{Joshua T.~Vogelstein\\ Department of Applied Mathematics \& Statistics,  Johns Hopkins University
        \and 
		R.~Jacob Vogelstein \\ National Security Technology Department, Johns Hopkins University Applied Physics Laboratory
		\and 
		Carey E.~Priebe \\ Department of Applied Mathematics \& Statistics,  Johns Hopkins University}

\date{}
\maketitle

%
%
%
%

	
\begin{abstract}
\noindent The``mind-brain supervenience'' conjecture suggests that all mental properties are derived from the physical properties of the brain. To address the question of whether the mind supervenes on the brain, we frame a supervenience hypothesis in rigorous statistical terms. Specifically, we propose a modified version of supervenience (called $\eps$-supervenience) that is amenable to experimental investigation and statistical analysis. To illustrate this approach, we perform a thought experiment that illustrates how the probabilistic theory of pattern recognition can be used to make a \emph{one}-sided determination of $\eps$-supervenience. The physical property of the brain employed in this analysis is the graph describing brain connectivity (i.e., the brain-graph or connectome). $\eps$-supervenience allows us to determine whether a particular mental property can be inferred from one's connectome to within any given positive misclassification rate, regardless of the relationship between the two. This may provide motivation for cross-disciplinary research between neuroscientists and statisticians.
\end{abstract}



\section*{Introduction}

\noindent Questions and assumptions about mind-brain supervenience go back at least as far as Plato's dialogues in circa 400 BCE \cite{Plato97}.  While there are many different notions of supervenience, we find Davidson's canonical description particularly illustrative \cite{Davidson70}:
\begin{quotation}
\noindent [mind-brain] supervenience might be taken to mean that there cannot be two events alike in all physical respects but differing in some mental respect, or that an object cannot alter in some mental respect without altering in some physical respect.
\end{quotation}
This philosophical conjecture has potentially widespread implications.  
For example, neural network theory and artificial intelligence often implicitly assume 
a local version mind-brain supervenience 
\cite{Haykin2008,Ripley2008}. Cognitive neuroscience similarly seems to operate under such assumptions
\cite{Gazzaniga2008}.  Philosophers continue to debate and refine notions of supervenience 
\cite{Kim2007}.  
Yet, to date, relatively scant attention has been paid to what might be empirically learned about supervenience.  

In this work we attempt to bridge the gap between philosophical conjecture and empirical investigations by casting supervenience in a probabilistic framework amenable to hypothesis testing. 
We then use the probabilistic theory of pattern recognition to determine
the limits of what one can and cannot learn about supervenience through data analysis.


\section*{Results}

\subsection*{Statistical supervenience: a definition} 

\noindent Let $\mc{M}=\{m_1, m_2, \ldots\}$ be the space of all possible minds and
let $\mc{B}=\{b_1,b_2,\ldots\}$ be the set of all possible brains.  $\mc{M}$ includes a mind for each possible collection of thoughts, memories, beliefs, etc.
$\mc{B}$ includes a brain for each possible position and momentum of all subatomic particles within the skull.  
Given these definitions, Davidson's conjecture may be concisely and formally stated thusly:  $m \neq m' \implies b \neq b'$, where $(m,b), (m',b') \in \mc{M} \times \mc{B}$ are mind-brain pairs.  This mind-brain supervenience relation does not imply an injective relation, a causal relation, or an identity relation (see Appendix for more details and some examples).  To facilitate both statistical analysis and empirical investigation, we convert this supervenience relation from a logical to a probabilistic relation.  

Let $F_{MB}$ indicate a joint distribution of minds and brains. Statistical supervenience can then be defined as follows:
\begin{defi}
\label{def:1} 
$\mM$ is said to \textit{statistically supervene} on $\mB$ for distribution $F=F_{MB}$, denoted $\mM \overset{S}{\sim}_F \mB$, if and only if $\PP[m \neq m' | b=b']=0$, or equivalently $\PP[m = m' | b = b']=1$. 
\end{defi}
\noindent Statistical supervenience is therefore a probabilistic relation on sets (which could be considered a generalization of correlation; see Appendix for details).

\subsection*{Statistical supervenience is equivalent to perfect classification accuracy} 
\label{sub:theoretical_results}

If minds statistically supervene on brains, 
then if two minds differ, there must be some brain-based difference to account for the mental difference.  This means that there must exist a deterministic function $g^*$ mapping each brain to its supervening mind. One could therefore, in principle, know this function. When the space of all possible minds is finite---that is, $|\mM| < \infty$---any function $g\from \mB \to \mM$ mapping from minds to brains is called a \emph{classifier}.  
Define misclassification rate, the probability that $g$ misclassifies $b$ under distribution $F=F_{MB}$,  as
\begin{align} \label{eq:1}
L_{F}(g) = \PP[g(B) \neq M] =  \sum_{(m,b) \in \mc{M} \times \mc{B}} \II\{g(b) \neq m\} \PP[B=b, M=m],	
\end{align}
where $\II\{\cdot\}$ denotes the indicator function taking value unity whenever its argument is true and zero otherwise.  The Bayes optimal classifier $g^*$ minimizes $L_{F}(g)$ over all classifiers:	
$g^* = \argmin_{g} L_{F}(g)$.
The \emph{Bayes error}, or Bayes risk, $L^*=L_{F}(g^*)$, is the minimum possible misclassification rate.

The primary result of casting supervenience in a statistical framework is the below theorem, which 
follows immediately from Definition \ref{def:1} and Eq. \eqref{eq:1}: 
\begin{thm}
\label{thm:1} 
$\mM \overset{S}{\sim}_{F} \mB \Leftrightarrow L^*= 0$.
\end{thm}


The above argument shows (for the first time to our knowledge) that statistical supervenience and zero Bayes error are equivalent. Statistical supervenience can therefore be thought of as a constraint on the possible distributions on minds and brains.  Specifically, let $\mc{F}$ indicate the set of all possible joint distributions on minds and brains, and let $\mc{F}_s = \{F_{MB} \in \mc{F}: L^*=0\}$ be the subset of distributions for which supervenience holds. Theorem \ref{thm:1} implies that $\mc{F}_s  \subsetneqq \mc{F}$.  Mind-brain supervenience is therefore an extremely restrictive assumption about the possible relationships between minds and brains.  It seems that such a restrictive assumption begs for empirical evaluation, vis-\'a-vis, for instance, a hypothesis test.

\subsection*{The non-existence of a viable statistical test for supervenience} 
\label{sub:subsection_name}


The above theorem implies that if we desire to know whether  minds supervene on brains, we can check whether $L^*=0$.  Unfortunately, $L^*$ is typically unknown.  Fortunately, we can approximate $L^*$ using training data.

Assume that training data $\mc{T}_n=\{(M_{1},B_{1}), \ldots, (M_{n},B_{n})\}$ are each sampled identically and independently (iid) from the true (but unknown) joint distribution $F=F_{MB}$.  Let $g_n$ be a classifier induced by the training data, $g_n:\mB \times (\mc{M} \times \mc{B})^n \mapsto \mM$.  The  misclassification rate of such a classifier is given by
\begin{align}
L_F(g_n)=\sum_{(m,b)  \in \mc{M}\times \mc{B}} \II\{g_n(b; \mc{T}_n) \neq m\} \PP[B=b,M=m],
\end{align}
which is a random variable due to the dependence on a randomly sampled training set $\mc{T}_n$.  
Calculating the expected misclassification rate $\EE[L_F(g_n)]$ is often intractable in practice because it requires a sum over all possible training sets.  Instead, expected misclassification rate can be approximated by ``hold-out'' error.  Let $\mc{H}_{n'}=\{(M_{n+1},B_{n+1}), \ldots, (M_{n+n'},B_{n+n'})\}$ be a set of $n'$ hold-out samples, each sampled iid from $F_{MB}$.  The hold-out approximation to the misclassification rate is given by
\begin{align}
\hL^{n'}_{F}(g_{n}) = \sum_{(M_i,B_i) \in \mc{H}_{n'}}\II \{g_{n}(B_i; \mc{T}_{n})\neq M_i\} \approx \EE[L_F(g_n)] \geq L^*. 
\end{align}
By definition of $g^*$, the expectation of $\hL^{n'}_F(g_n)$ (with respect to both $\mc{T}_n$ and $\mc{H}_{n'}$)  is greater than or equal to $L^*$ for any $g_n$ and all $n$.  Thus, we can construct a hypothesis test for $L^*$ using the surrogate $\hL^{n'}_F(g_n)$.  

A statistical test proceeds by 
specifying the allowable Type I error rate $\alpha>0$ and then
calculating a 
test statistic.
The $p$-value---the probability of rejecting the least favorable null hypothesis (the simple hypothesis within the potentially composite null which is closest to the boundary with the alternative hypothesis)---is the probability of observing a result at least as extreme as the observed.  In other words, the $p$-value is the cumulative distribution function of the test statistic evaluated at the observed test statistic with parameter given by the least favorable null distribution.
We reject if 
the $p$-value 
is less than $\alpha$.   A test is \emph{consistent} whenever its power  (the probability of rejecting the null when it is indeed false)  goes to unity as $n\conv \infty$ . For any statistical test, if the $p$-value converges in distribution to $\delta_0$ (point mass at zero), then whenever $\alpha >0$, power goes to unity. 

Based on the above considerations,
we might consider the following hypothesis test: $H_0: L^*>0$ and $H_A: L^*=0$; rejecting the null indicates that $\MsB$. Unfortunately, 
the alternative hypothesis lies on the boundary, so the $p$-value is always equal to unity \cite{Bickel2000}.  From this, Theorem \ref{thm:2} follows immediately:
\begin{thm} \label{thm:2}
	There does not exist a viable test of $\MsB$.
\end{thm}

In other words, we can \emph{never} reject $L^*>0$ in favor of supervenience, no matter how much data we obtain.  

\subsection*{Conditions for a consistent statistical test for $\eps$-supervenience} 

To proceed, therefore, we introduce a relaxed notion of supervenience: 
\begin{defi}
\label{def:2}
$\mM$ is said to $\eps$-\textit{supervene} on $\mB$ for distribution $F=F_{MB}$, denoted $\MeB$, if and only if $L^*< \eps$ for some $\eps>0$.
\end{defi}
\noindent 
Given this relaxation, consider the problem of testing for $\eps$-supervenience:
\begin{align*}
	H_0^{\eps}: L^* \geq \eps \\
	H_A^{\eps}: L^* < \eps.
\end{align*}
Let $\mh{n}= n' \hL^{n'}_{F}(g_n)$ be the \emph{test statistic}. 
The distribution of $\mh{n}$ is available under the least favorable null distribution. 
For the above hypothesis test,  
the $p$-value is therefore the binomial cumulative distribution function with parameter $\eps$; that is, $p$-value $=\mathbb{B}(\mh{n}; n', \eps)= \sum_{k \in [\mh{n}]_0}$Binomial$(k; n'; \eps)$, where
$[\mh{n}]_0=\{0,1,\ldots, \mh{n}\}$.  We reject whenever this $p$-value is less than $\alpha$; rejection implies that we are $100(1-\alpha$)\% confident that $\MeB$.   

 For the above $\eps$-supervenience statistical test, if $g_n \conv g^*$ as $n \conv \infty$, then $\hL^{n'}_F(g_n) \conv L^*$ as $n,n' \conv \infty$.  Thus, if $L^* < \eps$, 
power goes to unity.
The definition of $\eps$-supervenience therefore admits, for the first time to our knowledge, a viable statistical test of supervenience, given a specified $\eps$ and $\alpha$. Moreover, this test is consistent whenever $g_n$ converges to the Bayes classifier $g^*$.

\subsection*{The existence and construction of a consistent statistical test for $\eps$-supervenience} 
\label{sub:uc}

The above considerations indicate the existence of a consistent test for $\eps$-supervenience whenever the classifier used is consistent.  
To actually implement such a test, one must be able to (i) measure mind/brain pairs and (ii) have a consistent classifier $g_n$.  Unfortunately, we do not know how to measure the entirety of one's brain, much less one's mind. 
We therefore must restrict our interest to a mind/brain \emph{property} pair.  
A mind (mental) property might be a person's intelligence, psychological state, current thought, gender identity, etc.  A brain property might be the number of cells in a person's brain at some time $t$, or the collection of spike trains of all neurons in the brain during some time period $t$ to $t'$.  Regardless of the details of the specifications of the mental property and the brain property, given such specifications, one can assume a model, $\mc{F}$.  We desire a classifier $g_n$ that is guaranteed to be consistent, no matter which of the possible distributions $F_{MB} \in \mc{F}$ is the true distribution.  A classifier with such a property is called a \emph{universally consistent classifier}.  
Below, under a very general mind-brain model $\mc{F}$, we construct a universally consistent classifier. 


\begin{thex} \label{exp:1}
Let the physical property under consideration be brain connectivity structure, so $b$ is a brain-graph (``connectome'') with vertices representing neurons (or collections thereof) and edges representing synapses (or collections thereof). Further let $\mB$, the brain observation space, be the collection of all graphs on a given finite number of vertices, and let $\mc{M}$, the mental property observation space, be finite. Now, imagine collecting very large amounts of very accurate identically and independently sampled  brain-graph data and associated mental property indicators from $F_{MB}$. A $k_n$-nearest neighbor classifier using a Frobenius norm is universally consistent (see Methods for details). 
The existence of a universally consistent classifier guarantees that eventually (in $n,n'$) we will be able to conclude $\MeB$ for this mind-brain property pair, if indeed $\varepsilon$-supervenience holds. This logic holds for directed graphs or multigraphs or hypergraphs with discrete edge weights and vertex attributes, as well as unlabeled graphs (see \cite{VP11_unlabeled} for details). Furthermore, the proof holds for other matrix norms (which might speed up convergence and hence reduce the required $n$), and the regression scenario where $|\mM|$ is infinite (again, see Methods for details).  
\end{thex}
Thus, under the conditions stated in the above \emph{Gedankenexperiment}, universal consistency yields:
\begin{thm} \label{thm:3}
	$\MeB \implies \beta \conv 1$ as $n,n'\conv \infty$.
\end{thm}

Unfortunately, the rate of convergence of $L_{F}(g_n)$ to $L_{F}(g^*)$ depends on the (unknown) distribution $F=F_{MB}$ \cite{DGL96}. Furthermore, arbitrarily slow convergence theorems regarding the rate of convergence of $L_{F}(g_n)$ to $L_{F}(g^*)$ demonstrate that there is no universal $n,n'$ which will guarantee that the test has power greater than any specified target $\beta > \alpha$ \cite{Devroye83}. For this reason, the test outlined above can provide only a one-sided conclusion: if we reject we can be $100(1-\alpha)$\% confident that $\MeB$ holds, but we can never be confident in its negation; rather, it may be the case that the evidence in favor of $\MeB$ is insufficient 
because we simply have not yet collected enough data. 
This leads immediately to the following theorem:
\begin{thm} \label{thm:4}
For any target power $\beta_{min} > \alpha$, there is no universal $n,n'$ that guarantees $\beta \geq \beta_{min}$.
\end{thm}

Therefore, even $\eps$-supervenience does not satisfy Popper's falsifiability criterion \cite{Popper1959}.

\subsection*{The feasibility of a consistent statistical test for $\eps$-supervenience} 

Theorem \ref{thm:3} demonstrates the availability of a consistent test under certain restrictions.  Theorem \ref{thm:4}, however, demonstrates that convergence rates might be unbearably slow.  We therefore provide an illustrative example of the feasibility of such a test on synthetic data.

{\it Caenorhabditis elegans} is a species whose nervous system is believed to consist of the same $279$ labeled neurons for each organism \cite{Durbin87}. Moreover, these animals exhibit a rich behavioral repertoire that seemingly depends on circuit properties \cite{deBonoMaricq05}.  These findings motivate the use of C.~elegans for a synthetic data analysis \cite{GelmanShalizi11}.  Conducting such an experiment requires specifying a joint distribution $F_{MB}$ over brain-graphs and behaviors.  The joint distribution decomposes into the product of a class-conditional distribution (likelihood) and a prior, $F_{MB}=F_{B|M}F_M$. The prior  specifies the probability of any particular organism exhibiting the behavior.  The class-conditional distribution specifies the brain-graph distribution given that the organism does (or does not) exhibit the behavior. 

Let $A_{uv}$ be the number of chemical synapses between neuron $u$ and neuron $v$ according to \cite{VarshneyChklovskii09}.  Then, let $\mc{S}$ be the set of edges deemed responsible for odor-evoked behavior according to \cite{ChalasaniBargmann07}.  If odor-evoked behavior is supervenient on this signal subgraph $\mc{S}$, then the distribution of edges in $\mc{S}$ must differ between the two classes of odor evoked behavior \cite{VP11_sigsub}.  Let $E_{uv|j}$ denote the expected number of edges from vertex $v$ to vertex $u$ in class $j$.   For class $m_0$, let $E_{uv|0}=A_{uv}+\eta$,  where $\eta=0.05$ is a small noise parameter  (it is believed that the C.~elegans connectome is similar across organisms \cite{Durbin87}). For class $m_1$, let $E_{uv|1}=A_{uv}+z_{uv}$, where the signal parameter $z_{uv}=\eta$ for all edges not in $\mc{S}$, and $z_{uv}$ is uniformly sampled from $[-5,5]$ for all edges within $\mc{S}$. For both classes, let each edge be Poisson distributed, $F_{A_{uv}|M=m_j}$= Poisson$(E_{uv|j})$.

\begin{figure}[!ht]
\centering 
\includegraphics[width=.5\linewidth]{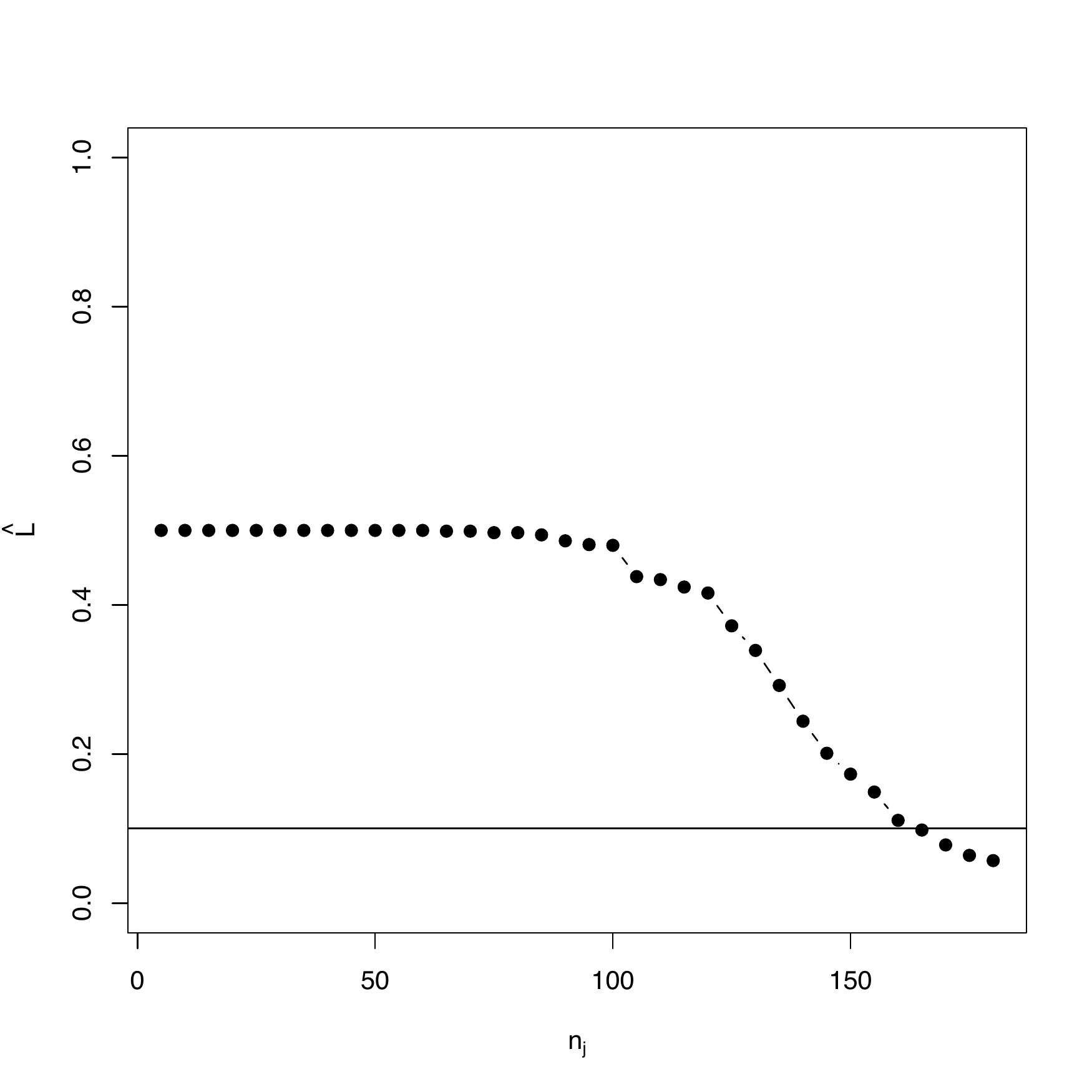}
\caption{C.~elegans graph classification simulation results.  The estimated hold-out misclassification rate $\hL^{n'}_{F}(g_{n})$  (with $n'=1000$ testing samples) 
is plotted as a function of class-conditional training sample size $n_j=n/2$, suggesting that for $\varepsilon=0.1$ we can determine that $\MeB$ holds with $99\%$ confidence with just a few hundred training samples generated from $F_{MB}$. Each dot depicts $\hL^{n'}_{F}(g_n)$ for some $n$; standard errors are $(\hL^{n'}_{F}(g_{n}) (1-\hL^{n'}_{F}(g_{n}))/n')^{1/2}$.  For example, at $n_j = 180$ we have $k_n = \lfloor\sqrt{8 n}\rfloor=53$ (where $\lfloor\cdot\rfloor$ indicates the floor operator), $\hL^{n'}_{F}(g_{n}) = 0.057$, and standard error less than $0.01$. We reject $H_0^{0.1}: L^* \geq 0.1$ at $\alpha=0.01$. Note that $L^* \approx 0$ for this simulation.}
\label{fig:1}
\end{figure}

We consider $k_n$-nearest neighbor classification of labeled multigraphs (directed, with loops) on 279 vertices, under Frobenius norm. The $k_n$-nearest neighbor classifier used here satisfies $k_n \rightarrow \infty$ as $n \rightarrow \infty$ and $k_n/n \rightarrow 0$ as $n \rightarrow \infty$, ensuring universal consistency. (Better classifiers can be constructed for the joint distribution $F_{MB}$ used here; however, we demand universal consistency.)  Figure \ref{fig:1} shows that for this simulation, rejecting $(\eps=0.1)$-supervenience at $\alpha=0.01$ requires only a few hundred training samples.


Importantly, conducting this experiment {\it in actu} is not beyond current technological limitations. 3D superresolution imaging \cite{Vaziri2008} combined with neurite tracing algorithms \cite{Helmstaedter2008,Mishchenko09,LuLichtman09} allow the collection of a C. elegans brain-graph within a day. Genetic manipulations, laser ablations, and training paradigms can each be used to obtain a non-wild type population for use as $M=m_1$ \cite{deBonoMaricq05}, and the class of each organism ($m_0$ vs.~$m_1$) can also be determined automatically \cite{Buckingham2008}.

\section*{Discussion}


This work makes the following contributions.  First, we define statistical supervenience based on Davidson's canonical statement (Definition \ref{def:1}).  This definition makes it apparent that supervenience implies the possibility of perfect classification (Theorem \ref{thm:1}).  
We then prove that there is no viable test against supervenience, so one can \emph{never} reject a null hypothesis in favor of supervenience, regardless of the amount of data (Theorem \ref{thm:2}).  This motivates the introduction of a relaxed notion called $\eps$-supervenience (Definition \ref{def:2}), against which consistent statistical tests are readily available. Under a very general brain-graph/mental property model (\emph{Gedankenexperiment} \ref{exp:1}),  a consistent statistical test against $\eps$-supervenience is always available no matter the true distribution $F_{MB}$ (Theorem \ref{thm:3}).  
In other words, the proposed test is guaranteed to reject the null whenever the null is false, given sufficient data, for any possible distribution governing mental property/brain property pairs. 

Alas, arbitrary slow convergence theorems demonstrate that there is no universal $n,n'$ for which convergence is guaranteed (Theorem \ref{thm:4}).  Thus, a failure to reject is ambiguous: even if the data satisfy the above assumptions, the failure to reject may be due to either (i) an insufficient amount of data or (ii) $\mc{M}$ may not be $\eps$-supervenient on $\mc{B}$.  Moreover, the data will not, in general, satisfy the above assumptions.  In addition to dependence (because each human does not exist in a vacuum), the mental property measurements will often be ``noisy'' (for example, accurately diagnosing psychiatric disorders is a sticky wicket \cite{Kessler2005}). 
%
Nonetheless, synthetic data analysis suggests that under somewhat realistic assumptions, convergence obtains with an amount of data one might conceivably collect (Figure \ref{fig:1} and ensuing discussion).

Thus, given measurements of mental and brain properties that we believe reflect the properties of interest, and given a sufficient amount of data satisfying the independent and identically sampled assumption, a rejection of $H_0^{\eps}: L^*\geq \eps$ in favor of $\MeB$ entails that we are $100(1-a)\%$ confident that the mental property under investigation is $\eps$-supervenient on the brain property under investigation.  Unfortunately, failure to reject is more ambiguous. 

Interestingly, much of contemporary research in neuroscience and cognitive science could be cast as mind-brain supervenience investigations.  Specifically, searches for ``engrams'' of  memory traces \cite{Lashley50} or ``neural correlates''  of various behaviors or mental properties (for example, consciousness \cite{Koch2010}), may be more aptly called searches for the ``neural supervenia'' of such properties.
Letting the brain property be a brain-graph is perhaps especially pertinent in light of the advent of ``connectomics'' \cite{SpornsKotter05,Hagmann05}, a field devoted to estimating whole organism brain-graphs and relating them to function.  Testing supervenience of various mental properties on these brain-graphs will perhaps therefore become increasingly compelling; the framework developed herein could be fundamental to these investigations.  
For example, questions about whether connectivity structure alone is sufficient to explain a particular mental property is one possible mind-brain $\eps$-supervenience investigation.  The above synthetic data analysis demonstrates the feasibility of $\eps$-supervenience on small brain-graphs.
Similar supervenience tests on larger animals (such as humans) will potentially benefit from either higher-throughput imaging modalities \cite{HayworthLichtman06, Bock2011}, more coarse brain-graphs \cite{PalmAmunts10,Johansen-Berg2009}, or both.



\section*{Methods}
\label{sec:methods}


The $1$-nearest neighbor ($1$-NN) classifier works as follows.  Compute the distance between the test brain  $b$ and all $n$ training brains, $d_i=d(b,b_i)$ for all $i \in [n]$, where $[n]=1,2,\ldots, n$.  Then, sort these distances, $d_{(1)} < d_{(2)} < \ldots < d_{(n)}$, and consider their corresponding minds, $m_{(1)}, m_{(2)}, \ldots, m_{(n)}$, where parenthetical indices indicate rank order among $\{d_i\}_{i\in[n]}$.  
The $1$-NN algorithm predicts that the unobserved mind is of the same class as the closest brain's class: $\mh{m}=m_{(1)}$.  The $k_n$ nearest neighbor is a straightforward generalization of this approach.  It says that the test mind is in the same class as whichever class is the plurality class among the $k_n$ nearest neighbors, $\mh{m}=\argmax_{m'}\II\{\sum_{i=1}^{k_n} m_{(i)}=m'\}$.  Given a particular choice of $k_n$ (the number of nearest neighbors to consider) and a choice of $d(\cdot,\cdot)$ (the distance metric used to compare the test datum and training data), one has a relatively simple and intuitive algorithm.  

Let $g_n$ be the $k_n$ nearest neighbor ($k_n$NN) classifier when there are $n$ training samples.  
A collection of such classifiers $\{g_n\}$,  with $k_n$ increasing with $n$, is called a classifier sequence.  
A universally consistent classifier sequence is any classifier sequence that is guaranteed to converge to the Bayes optimal classifier regardless of the true distribution from which the data were sampled; that is, a universally consistent classifier sequence satisfies $L_F(g_n) \conv L_F(g^*)$ as $n \conv \infty$ for all $F_{MB}$. In the main text, we refer to the whole sequence as a classifier.

The $k_n$NN classifier is consistent if (i) $k_n \conv \infty$ as $n \conv \infty$ and (ii) $k_n/n \conv 0$ as $n\conv\infty$ \cite{Stone1977}. In Stone's original proof \cite{Stone1977}, $b$ was assumed to be a $q$-dimensional vector, and the $L_2$ norm ($d(b,b')=\sum_{j=1}^q (b_j-b_j')^2$, where $j$ indexes elements of the $q$-dimensional vector) was shown to satisfy the constraints on a distance metric for this collection of classifiers to be universally consistent.  Later, others extended these results to apply to any $L_p$ norm \cite{DGL96}.  When brain-graphs are represented by their adjacency matrices, one can stack the columns of the adjacency matrices, effectively embedding graphs into a vector space, in which case Stone's theorem applies.  Stone's original proof also applied to the scenario when $|\mc{M}|$ was infinite, resulting in a universally consistent regression algorithm as well.

Note that the above extension of Stone's original theorem to the graph domain implicitly assumed that vertices were labeled, such that elements of the adjacency matrices could easily be compared across graphs.  In theory, when vertices are unlabeled, one could first map each graph to a quotient space invariant to isomorphisms, and then proceed as before.  Unfortunately, 
there is no known polynomial time complexity algorithm for graph isomorphism
\cite{GareyJohnson79}, so in practice, dealing with unlabeled vertices will likely be computationally challenging \cite{VP11_unlabeled}.






\newpage
\appendix
\section*{Appendix}

\begin{figure}[h!tbp]
	\centering
		\includegraphics[width=1\linewidth]{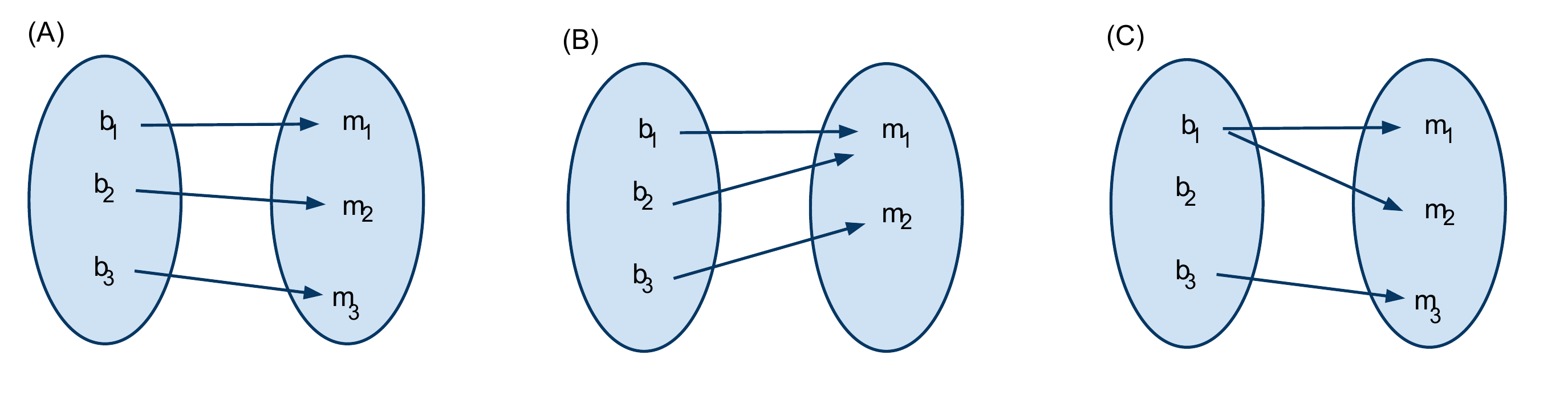}
	\caption{Possible relations between minds and brains.  (A) Minds supervene on brains, and it so happens that there is a bijective relation from brains to minds.  (B) Minds supervene on brains, and it so happens that there is a surjective (a.k.a., onto) relation from brains to minds, but not a bijective one. (C) Minds are \emph{not} supervenient on brains, because two different minds supervene on the same brain.}
	\label{fig:rel}
\end{figure}

In this appendix we compare supervenience to several other relations on sets (see Figure \ref{fig:rel}).



First, a supervenient relation does not imply an injective relation.  An injective relation is any relation that preserves distinctness.  Thus if minds are injective on brains, then $b\neq b' \implies m \neq m'$ (note that the directionality of the implication has been switched relative to supervenience). However, it might be the case that a brain could change without the mind changing.  Consider the case that a single subatomic particle shifts its position by a Plank length, changing brain state from $b$ to $b'$.  It is possible (likely?) that the mental state supervening on brain state $b$ remains $m$, even after $b$ changes to $b'$.  In such a scenario, the mind might still supervene on the brain, but the relation from brains to minds is not injective. This argument also shows that supervenience is not necessarily a \emph{symmetric} relation.  Minds supervening on brains does not imply that brains supervene on minds.

Second,  supervenience does not imply causality. 
For instance, consider an analogy where $M$ and $B$ correspond to two coins being flipped, each possibly landing on heads or tails.  Further assume that every time one lands on heads so does the other, and every time one lands on tails, so do the other. This implies that $M$ supervenes on $B$, but assumes nothing about whether $M$ causes $B$, or $B$ causes $M$, or some exogenous force causes both.  

Third, supervenience does not imply identity.  The above example with the two coins demonstrates this, as the two coins are not the same thing, even if one has perfect information about the other.  

What supervenience does imply, however, is the following.   Imagine finding two unequal minds.  If $\MsB$, then the brains on which those two minds supervene must be different.  In other words, there cannot be two unequal minds, either of which could supervene on a single brain.  Figure \ref{fig:rel} shows several possible relations between the sets of minds and brains.

Note that statistical supervenience is distinct from statistical correlation.  \emph{Statistical correlation} between brain states and mental states is defined as $\rho_{MB}=\EE[(B-\mu_B)(M-\mu_M)]/(\sig_B \sig_M)$, where $\mu_X$ and $\sig_X$ are the mean and variance of $X$, and $\EE[X]$ is the expected value of $X$.  If $\rho_{MB}=1$, then both $\MsB$ and $\mB \overset{S}{{\sim}}_{F} \mM$. Thus, perfect correlation implies supervenience, but supervenience does not imply correlation.  In fact, supervenience may be thought of as a generalization of correlation which incorporates directionality, can be applied to arbitrary valued random variables (such as mental or brain properties), can depend on any moment of a distribution (not just the first two).

\clearpage
\bibliography{/Users/jovo/Research/latex/library}

\begin{thebibliography}{10}

\bibitem{Plato97}
Plato.
\newblock {\em {Plato: complete works}}.
\newblock Hackett Pub Co,  (1997).

\bibitem{Davidson70}
Davidson, D.
\newblock {\em {Experience and Theory}}, chapter Mental Eve.
\newblock Duckworth (1970).

\bibitem{Haykin2008}
Haykin, S.
\newblock {\em {Neural Networks and Learning Machines}}.
\newblock Prentice Hall, 3rd edition,  (2008).

\bibitem{Ripley2008}
Ripley, B.~D.
\newblock {\em {Pattern Recognition and Neural Networks}}.
\newblock Cambridge University Press,  (2008).

\bibitem{Gazzaniga2008}
Gazzaniga, M.~S., Ivry, R.~B., and Mangun, G.~R.
\newblock {\em {Cognitive Neuroscience: The Biology of the Mind (Third
  Edition)}}.
\newblock W. W. Norton \& Company,  (2008).

\bibitem{Kim2007}
Kim, J.
\newblock {\em {Physicalism, or Something Near Enough (Princeton Monographs in
  Philosophy)}}.
\newblock Princeton University Press,  (2007).

\bibitem{Bickel2000}
Bickel, P.~J. and Doksum, K.~A.
\newblock {\em {Mathematical Statistics: Basic Ideas and Selected Topics, Vol I
  (2nd Edition)}}.
\newblock Prentice Hall,  (2000).

\bibitem{VP11_unlabeled}
Vogelstein, J.~T. and Priebe, C.~E.
\newblock {\em Submitted for publication}{ \bf } (2011).

\bibitem{DGL96}
Devroye, L., Gy\"{o}rfi, L., and Lugosi, G.
\newblock {\em {A Probabilistic Theory of Pattern Recognition}}.
\newblock Springer,  (1996).

\bibitem{Devroye83}
Devroye, L.
\newblock {\em Utilitas Mathematica}{ \bf 483}, 475--483 (1983).

\bibitem{Popper1959}
Popper, K.~R.
\newblock {\em {The logic of scientific discovery}}.
\newblock Routledge,  (1959).

\bibitem{Durbin87}
Durbin, R.~M.
\newblock {\em {Studies on the Development and Organisation of the Nervous
  System of Caenorhabditis elegans}}.
\newblock PhD thesis, University of Cambridge,  (1987).

\bibitem{deBonoMaricq05}
de~Bono, M. and Maricq, A.~V.
\newblock {\em Annu Rev Neurosci}{ \bf 28}, 451--501 (2005).

\bibitem{GelmanShalizi11}
Gelman, A. and Shalizi, C.~R.
\newblock {\em Submitted for publication}{ \bf }, 1--36 (2011).

\bibitem{VarshneyChklovskii09}
Varshney, L.~R., Chen, B.~L., Paniagua, E., Hall, D.~H., Chklovskii, D.~B.,
  Spring, C., and Farm, J.
\newblock {\em World Wide Web Internet And Web Information Systems}{ \bf },
  1--41.

\bibitem{ChalasaniBargmann07}
Chalasani, S.~H., Chronis, N., Tsunozaki, M., Gray, J.~M., Ramot, D., Goodman,
  M.~B., and Bargmann, C.~I.
\newblock {\em Nature}{ \bf 450}(7166), 63--70 November  (2007).

\bibitem{VP11_sigsub}
Vogelstein, J.~T., Gray, W.~R., Vogelstein, R.~J., and Priebe, C.~E.
\newblock {\em Submitted for publication}{ \bf } (2011).

\bibitem{Vaziri2008}
Vaziri, A., Tang, J., Shroff, H., and Shank, C.~V.
\newblock {\em Proceedings of the National Academy of Sciences of the United
  States of America}{ \bf 105}(51), 20221--6 December  (2008).

\bibitem{Helmstaedter2008}
Helmstaedter, M., Briggman, K.~L., and Denk, W.
\newblock {\em Current opinion in neurobiology}{ \bf 18}(6), 633--41 December
  (2008).

\bibitem{Mishchenko09}
Mishchenko, Y.
\newblock {\em J Neurosci Methods}{ \bf 176}(2), 276--289 January  (2009).

\bibitem{LuLichtman09}
Lu, J., Fiala, J.~C., and Lichtman, J.~W.
\newblock {\em PLoS ONE}{ \bf 4}(5), e5655 (2009).

\bibitem{Buckingham2008}
Buckingham, S.~D. and Sattelle, D.~B.
\newblock {\em Invertebrate neuroscience : IN}{ \bf 8}(3), 121--31 September
  (2008).

\bibitem{Kessler2005}
Kessler, R.~C., Berglund, P., Demler, O., Jin, R., Merikangas, K.~R., and
  Walters, E.~E.
\newblock {\em Archives of general psychiatry}{ \bf 62}(6), 593--602 June
  (2005).

\bibitem{Lashley50}
Lashley, K.~S.
\newblock {\em Symposia of the society for experimental biology}{ \bf
  4}(454-482), 30 (1950).

\bibitem{Koch2010}
Koch, C.
\newblock {\em {The Quest for Consciousness}}.
\newblock Roberts and Company Publishers,  (2010).

\bibitem{SpornsKotter05}
Sporns, O., Tononi, G., and Kotter, R.
\newblock {\em PLoS Computational Biology}{ \bf 1}(4), e42 (2005).

\bibitem{Hagmann05}
Hagmann, P.
\newblock {\em {From diffusion MRI to brain connectomics}}.
\newblock PhD thesis, Institut de traitement des signaux,  (2005).

\bibitem{HayworthLichtman06}
Hayworth, K.~J., Kasthuri, N., Schalek, R., Lichtman, J.~W., Program, N.,
  Angeles, L., and Biology, C.
\newblock {\em World}{ \bf 12}(Supp 2), 86--87 (2006).

\bibitem{Bock2011}
Bock, D.~D., Lee, W.-C.~A., Kerlin, A.~M., Andermann, M.~L., Hood, G., Wetzel,
  A.~W., Yurgenson, S., Soucy, E.~R., Kim, H.~S., and Reid, R.~C.
\newblock {\em Nature}{ \bf 471}(7337), 177--182 March  (2011).

\bibitem{PalmAmunts10}
Palm, C., Axer, M., Gr\"{a}\ss~el, D., Dammers, J., Lindemeyer, J., Zilles, K.,
  Pietrzyk, U., and Amunts, K.
\newblock {\em Frontiers in Human Neuroscience}{ \bf 4} (2010).

\bibitem{Johansen-Berg2009}
Johansen-Berg, H. and Behrens, T.~E.
\newblock {\em {Diffusion MRI: From quantitative measurement to in-vivo
  neuroanatomy}}.
\newblock Academic Press,  (2009).

\bibitem{Stone1977}
Stone, C.~J.
\newblock {\em The Annals of Statistics}{ \bf 5}(4), 595--620 July  (1977).

\bibitem{GareyJohnson79}
Garey, M.~R. and Johnson, D.~S.
\newblock {\em {Computers and intractability. A guide to the theory of
  NP-completeness. A Series of Books in the Mathematical Sciences}}.
\newblock WH Freeman and Company, San Francisco, Calif,  (1979).

\end{thebibliography}
\bibliographystyle{nature}

\section*{Acknowledgments}

The authors would like to acknowledge helpful discussions with J.~Lande, B.~Vogelstein, S.~Seung, and two helpful referees.

\section*{Author Contributions}

JTV, RJV, and CEP conceived of the manuscript.  JTV and CEP wrote it.  CEP ran the experiment.

\section*{Additional Information}

The authors have no competing financial interests to declare.

\end{document}